\documentclass[aps,prx,twocolumn,showpacs,superscriptaddress]{revtex4-1}

\usepackage{latexsym}
\usepackage{ifpdf}
\usepackage{graphicx}
\usepackage{epsfig}
\usepackage{amsmath, amssymb, bbm, mathrsfs}
\usepackage{multirow}
\usepackage{ulem}
\usepackage{fancyhdr}
\usepackage{color}
\usepackage{bm}
\usepackage{eurosym}
\usepackage{subfigure}
\usepackage{comment}
\usepackage{scrhack}
\usepackage{physics}
\usepackage{balance}
\usepackage{float}
\usepackage{flushend}
\usepackage{textcomp}
\usepackage{setspace}   

\usepackage[unicode=true,pdfusetitle,
 bookmarks=false,
 breaklinks=true,pdfborder={0 0 0},pdfborderstyle={},backref=false,colorlinks=true]
 {hyperref}
\hypersetup{
 colorlinks,linkcolor=red,citecolor=blue}

\makeatletter
\newcommand*{\rom}[1]{\expandafter\@slowromancap\romannumeral #1@}
\makeatother
\usepackage{romannum}

\begin{document}

\title{Ground-state Pulsed Cavity Electro-optics for Microwave-to-optical Conversion}

\author{Wei Fu}
\thanks{These authors contributed equally to this work.}
\affiliation{Department of Electrical Engineering, Yale University, New Haven, Connecticut 06520, USA}

\author{Mingrui Xu}
\thanks{These authors contributed equally to this work.}
\affiliation{Department of Electrical Engineering, Yale University, New Haven, Connecticut 06520, USA}

\author{Xianwen Liu}
\affiliation{Department of Electrical Engineering, Yale University, New Haven, Connecticut 06520, USA}

\author{Chang-Ling Zou}
\affiliation{Department of Electrical Engineering, Yale University, New Haven, Connecticut 06520, USA}


\author{Changchun Zhong}
\affiliation{Pritzker School of Molecular Engineering, University of Chicago, Illinois 60637, USA}

\author{Xu Han}
\affiliation{Department of Electrical Engineering, Yale University, New Haven, Connecticut 06520, USA}

\author{Mohan Shen}
\affiliation{Department of Electrical Engineering, Yale University, New Haven, Connecticut 06520, USA}

\author{Yuntao Xu}
\affiliation{Department of Electrical Engineering, Yale University, New Haven, Connecticut 06520, USA}

\author{Risheng Cheng}
\affiliation{Department of Electrical Engineering, Yale University, New Haven, Connecticut 06520, USA}

\author{Sihao Wang}
\affiliation{Department of Electrical Engineering, Yale University, New Haven, Connecticut 06520, USA}

\author{Liang Jiang}
\affiliation{Pritzker School of Molecular Engineering, University of Chicago, Illinois 60637, USA}
\affiliation{Yale Quantum Institute, Yale University, New Haven, Connecticut 06520, USA}

\author{Hong X. Tang}
\email{hong.tang@yale.edu}
\affiliation{Department of Electrical Engineering, Yale University, New Haven, Connecticut 06520, USA}
\affiliation{Yale Quantum Institute, Yale University, New Haven, Connecticut 06520, USA}


\begin{abstract}
\textbf{Abstract:} 
In the development of quantum microwave-to-optical (MO) converters, excessive noise induced by the parametric optical drive remains a major challenge at milli-Kelvin temperatures.  Here we study the extraneous noise added to an electro-optic transducer in its quantum ground state under an intense pulsed optical excitation. The integrated electro-optical transducer leverages the inherent Pockels effect of aluminum nitride microrings, flip-chip bonded to a superconducting resonator. Applying a pulsed optical drive with peak power exceeding the cooling power of the dilution refrigerator at its base temperature, we observe efficient bi-directional MO conversion, with near-ground state microwave thermal excitation ($\bar{n}_\mathrm{e}=0.09\pm0.06$).
Time evolution study reveals that the residual thermal excitation is dominated by the superconductor absorption of stray light scattered off the chip-fiber interface. 
Our results shed light on suppressing microwave noise in a cavity electro-optic system under intense optical drive, which is an essential step towards quantum state transduction between microwave and optical frequencies.

\end{abstract}

\maketitle

Hybrid superconducting-photonics systems operated in cryogenic conditions has drawn wide attention in the past decade, in pursuit of quantum microwave-to-optical (MO) conversion, microwave-optical quantum entanglement distribution, etc, as a building block of the envisioned future quantum network \cite{network1,network2,entanglement_cczhong,network3,network4}. 
Among many promising platforms including spin ensemble \cite{spin1,spin2}, erbium-doped crystal \cite{rare-earth-doped1,rare-earth-doped2}, ferromagnetic magnons \cite{magnon1,magnon2}, and electro-opto-mehcanics \cite{optomech_cleland,optomech_nist,optomech_nist_feedforward_natphy_2018,optomech_simon,optomech_changling,optomech_xu,optomech_painter}, integrated cavity electro-optical system \cite{eo1_Tsang_PRA_theory_2011,eo2_Kippenberg_PRA_theory_2016,eo3_Fink_Optica_2016,eo8_fan2018superconducting,eo4_loncar_doublering_pra_2017,eo5_LN_amir_2020,eo6_LN_loncar_2020,eo7_groundstate_fink_2020} stands out as a particularly interesting device architecture. 
Harnessing the Pockels effect, integrated electro-optical platforms offer advantages including tunability, scalability and high power handling capability because the device does not rely on free-standing structures and has a relatively large mode volume.
A recent work based on aluminum nitride (AlN) cavity electro-optics has demonstrated electromagnetically induced transparency and a state-of-art conversion efficiency of 2\% at 1.7\,K \cite{eo8_fan2018superconducting}. 
Orders of magnitudes of improvement in electro-optical coupling strength becomes possible with the recent advance in material platforms including lithium niobite \cite{LN_modulator,eo5_LN_amir_2020,eo4_loncar_doublering_pra_2017} and barium titanate \cite{BTO_modulator}.

For quantum transduction applications, it is crucial to obtain both high electro-optical interaction rates and low thermal occupations, which is assisted by integrating an intense optical parametric drive into device operation at the milli-Kelvin temperature \cite{Heating_PRA_Painter_2014,Heating_pulsed_PRX_Painter_2015,eo7_groundstate_fink_2020,optomech_nist,optomech_nist_feedforward_natphy_2018}. 
This is a particularly challenging task because the optical drive, of which the peak power might exceed the dilution refrigerator cooling power at the base temperature, induces extra microwave thermal excitations through photon absorption by dielectric and superconductor \cite{2000_apl_quasiparticles_NbN,Nature_2003_SC_photondetector_mazin,PRL2011NbNquasipartical,APL2013TiNquasipartical}. 
Promising progress has been made by recent demonstrations of MO conversion with ground-state electro-optomechanics \cite{optomech_simon,optomech_painter} and bulk electro-optics \cite{eo7_groundstate_fink_2020}. 
However, to achieve close-to-unitary conversion efficiency, which is required for quantum state transduction \cite{FOM_QUANTUMTRANSDUCER_2020}, implementation of stronger parametric optical drives in a microresonator system seems inevitable.
Therefore, a systematic understanding of thermal dynamics of the microwave resonance in a hybrid superconducting-photonic system, which may shed light on further suppressing the microwave noise in presence of a strong optical drive, is highly anticipated.

\begin{figure*}
\begin{centering}
\includegraphics{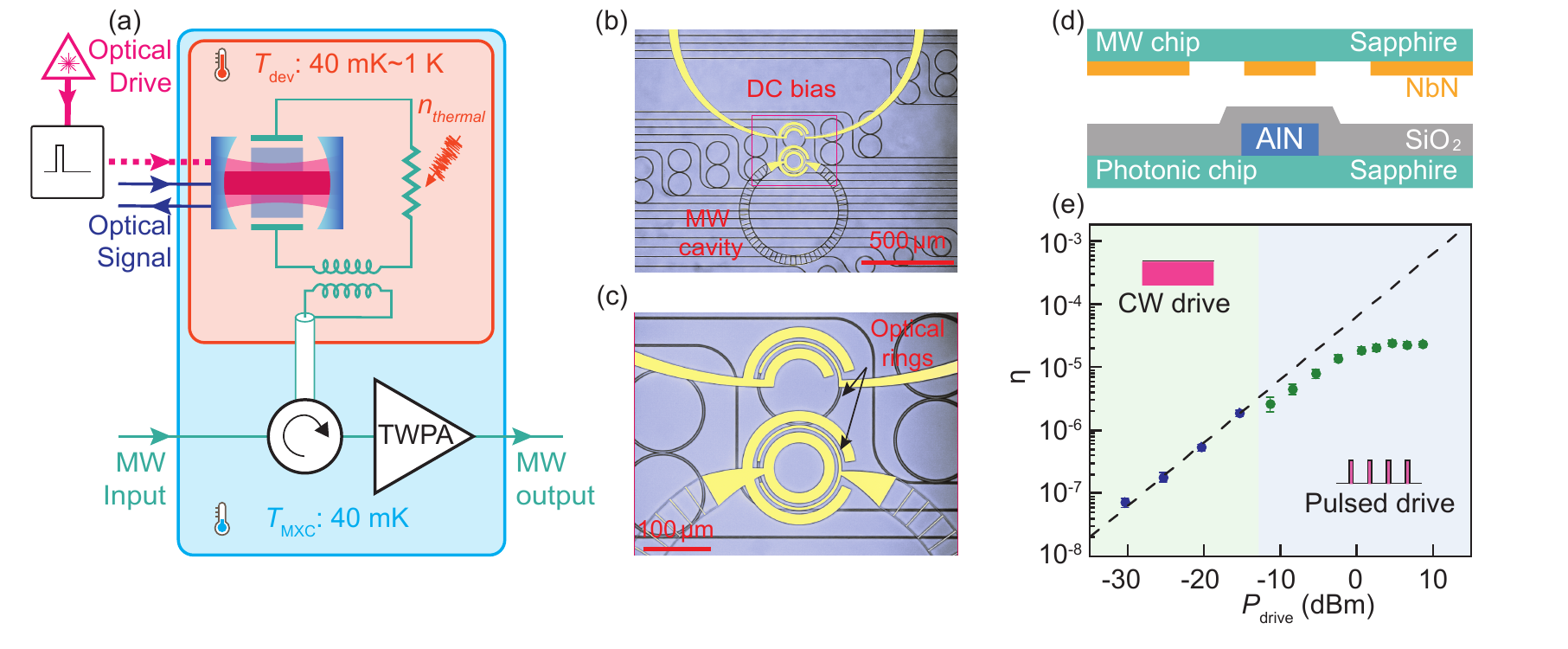}
\par\end{centering}
\caption{(a) A schematic of the experiment: an electro-optical (EO) transducer is installed in a milli-Kelvin environment. Optical drive is sent to the device to enhance parametric EO coupling but also inevitably introduces extraneous noises. A travelling wave amplifier (TWPA) is used to provide low-added noise readout of the microwave signal and noise from its output port. 
(b) and (c) are images of the EO transducer device. (d) Schematic of the flip-chip design. (e) Calibrated microwave-to-optical conversion efficiency as a function of optical drive power in the photonic waveguide. The blue dots are data measured with the continuous wave (CW) scheme; the olive dots are data measured with the pulsed scheme, as illustrated by the inset cartoons. Error bars represent the fitting error. The dashed line shows theoretically predicted conversion efficiency extrapolated from the CW regime.}
\label{fig1}
\end{figure*}

In this work, we study near-ground-state microwave thermal excitation induced by strong pulsed optical drives in an integrated electro-optical device mounted in a dilution refrigerator at milli-Kelvin. The microwave thermal noise from the device microwave port is calibrated through precise noise thermometry assisted by an ultra-low-noise travelling wave amplifier \cite{TWPA}, as illustrated in Fig.\,\ref{fig1}(a). 
By comparing the thermal occupancy of the microwave mode under continuous wave (CW) optical drives and pulsed optical drives, we found that the pulsed scheme allows around 30 dB higher peak pump power, while maintaining the same level of thermal excitations. 
In the presence of a pulsed optical drive with peak power of $-11.4\,\mathrm{dBm}$ (72\,$\mathrm{\mu W}$) in the waveguide, which exceeds the cooling power $50\,\mathrm{\mu W}$ of the dilution refrigerator at the base temperature around 40\,mK, the microwave mode occupancy is observed to be $0.09\pm0.06$, which corresponds to $92\pm5\%$ ground state probability.
Further study of the heating mechanisms suggests that the superconductor absorption of stray light scattered from the parametric drive, which creates an effective thermal bath with a response time much faster than the $200\,\mathrm{ns}$ measurement time constant, 
is a major limit to achieving microwave ground state when applying stronger optical drives.

\textit{Device design and implementation.---}The electro-optical transducer studied in this work utilizes a triple-resonant scheme: a microwave mode couples to two optical modes via the Pockels effect. In our previous implementations \cite{eo8_fan2018superconducting}, a pair of hybrid, dispersion-engineered transverse electric (TE)/transverse magnetic (TM) modes are utilized to satisfy electro-optical phase matching. Here we adopt a double-ring structure \cite{eo4_loncar_doublering_pra_2017,eo5_LN_amir_2020,eo6_LN_loncar_2020} to allow on-chip tuning of optical modal splitting by applying a DC voltage through integrated DC bias electrode. Figures \ref{fig1}(b) and \ref{fig1}(c) present images of the transducer. The double-ring structure supports hybridized symmetric and anti-symmetric TE optical modes for optical drive ($a$ mode) and signal ($b$ mode).  A tunable ouroboros LC resonator \cite{ouroboros} made of niobium nitride (NbN) supports the microwave mode ($c$ mode), where the capacitor is composed of three concentric ground-signal-ground electrodes and generates electric field perpendicular to the surface of the bottom ring in the proximity of the middle electrode. 
The inductor part consists wires with high kinetic inductance which provides resonant frequency tunability. 

In this electro-optical system, the interaction Hamiltonian is $H_{\mathrm{int}}=\hbar g_{\mathrm{o}}\left(ab^{\dagger}c+ a^{\dagger}bc^{\dagger} \right)$, where $a$, $b$, and $c$ are the annihilation operators for the optical drive mode, optical signal mode, and microwave modes, respectively, and $g_{\mathrm{o}}$ is the vacuum electro-optical coupling rate due to Pockels effect.
Pumping the lower frequency optical mode $a$ enables coherent conversion between the optical mode $b$ and the microwave mode $c$. When the two optical modes are separated by the microwave mode's frequency, we have the peak conversion efficiency \cite{eo8_fan2018superconducting}: 
\begin{equation}
\label{efficiency}
   \eta=\eta_\mathrm{c}\eta_\mathrm{b}\frac{4C}{\left(1+C\right)^2}=\frac{\kappa_{\mathrm{c,ex}}}{\kappa_{\mathrm{c}}}\frac{\kappa_{\mathrm{b,ex}}}{\kappa_{\mathrm{b}}}\frac{4C}{\left(1+C\right)^2},
\end{equation}
where $\eta_\mathrm{b}$ ($\eta_\mathrm{c}$), $\kappa_{\mathrm{b,ex}}$ ($\kappa_{\mathrm{c,ex}}$), and $\kappa_{\mathrm{b}}$ ($\kappa_{\mathrm{c}}$) are the extraction ratio, external loss rate, and total loss rate of the optical mode $b$ (microwave mode $c$). The electro-optical cooperativity $C=\frac{4n_\mathrm{drive} g_{\mathrm{o}}^2}{\kappa_\mathrm{b}\kappa_\mathrm{c}}$ is resonantly enhanced by the optical drive with an intracavity drive photon number $n_\mathrm{drive}$.


The microwave and photonic chips are fabricated independently and then assembled through flip-chip bonding, as illustrated in Fig. \ref{fig1}(d).
The photonic circuit is fabricated from a 1-$\mu$m thick crystalline AlN layer grown on a sapphire substrate, and is then capped by $1.5\,\mathrm{\mu m}$-thick silicon dioxide cladding prior to the waveguide facet cleavage. The microwave resonator and the DC bias electrode are patterned from a 50-nm NbN thin film deposited on a sapphire substrate. To maximize the overlap between the microwave and optical modes for a larger $g_{\mathrm{o}}$, the two chips are flip-chip assembled via a home-built bonding station with an in-plane alignment error less than 1\,$\mu$m. Fibers are glued to the side of the chip to send and collect light. The total insertion loss between the input and output fiber connectors outside the refrigerator is calibrated to be around 12\,dB. A coaxial cable terminated with a hoop antenna is employed to inductively couple with the microwave mode for input and readout. To shield external radiation and provide thermal anchor, we house the device under test (DUT) in a copper box, which is installed in the MXC of a dilution refrigerator.\par

When a resonant optical drive is sent to the device through the fiber, a large portion of the input optical power, which is around 50\% based on a room temperature calibration, is not coupled to the waveguide and scattered off the chip-fiber interface. These uncoupled photons, when absorbed by the device assembly, lead to heating of the chip as well as the packaging.
The coupled light propagates along the waveguide, and eventually dissipates in the microring, inducing local heating in the cavity. 
Heat generated in the device is dissipated through thermal contact between sapphire substrate and the copper housing, which is well thermalized to the MXC.

\textit{Device performance under pulsed drive.---}
To parametrically enhance the electro-optical coupling without excessively heating up the device as well as the MXC, we implement a pulsed drive scheme. 
As illustrated in Fig. \ref{fig1}(a), a CW laser from a tunable laser diode is modulated by an electrical pulse via an acousto-optic modulator (AOM), of which rise/fall time is 35\,ns, before being sent into the device.
Upon decreasing the duty cycle of the pulse signal, we essentially reduce the heat load to the device while maintain the peak intracavity photon number. 
In the following described experiments, we use pulses with 0.2\% duty cycle and 1\,ms period.
Thanks to the large extinction of the AOM, the optical drive power on-off ratio is over $50\,\mathrm{dB}$.
As a result, the heating caused by the leaked optical drive while off-duty can be neglected.
With such a pulsed drive scheme, we can send a laser drive with 0\,dBm peak power while the MXC temperature only rises 1\,mK from the base temperature.

By measuring the spectra of the complete tranducer's scattering matrix, we calibrated the conversion efficiency with both CW and pulsed drive of different powers.
In Fig. \ref{fig1}(e), blue dots represent efficiencies measured with CW drives, and olive dots represent efficiencies measured with pulsed drives, as illustrated by the inset cartoons.
The dashed line shows theoretically predicted conversion efficiency based on Eq.\,\ref{efficiency} with the coefficient $g_\mathrm{o}$\,=\,42\,$\mathrm{Hz}$ and and  microwave $Q$ extracted from measurements with a $-25.4$\,dBm CW drive.
The highest efficiency observed is $2.4\cross10^{-5}$ when pumping with 4.6\,dBm pulsed drive, which corresponds to an internal efficiency of $0.12\%$.
Slightly lower conversion efficiency with low power ($<$\,0\,dBm) pulsed drive, compared with the CW extrapolation, can be explained by the $Q$ drop of the microwave mode, because of the increased intrinsic decay rate induced by less saturated two-level systems at reduced ambient temperatures \cite{PhysRevApplied2019HighKIWire,apl2008TLS} and quasi-particle generated by superconductor absorption of optical photons \cite{2000_apl_quasiparticles_NbN,Nature_2003_SC_photondetector_mazin,PRL2011NbNquasipartical,APL2013TiNquasipartical}. In the higher power regime ($>$\,0\,dBm), the efficiencies deviate more from the theoretical value, possibly due to parasitic nonlinear effects in the ring resonator such as thermal-optic effect \cite{thermal_vahala}.

\begin{figure}[h]
\includegraphics{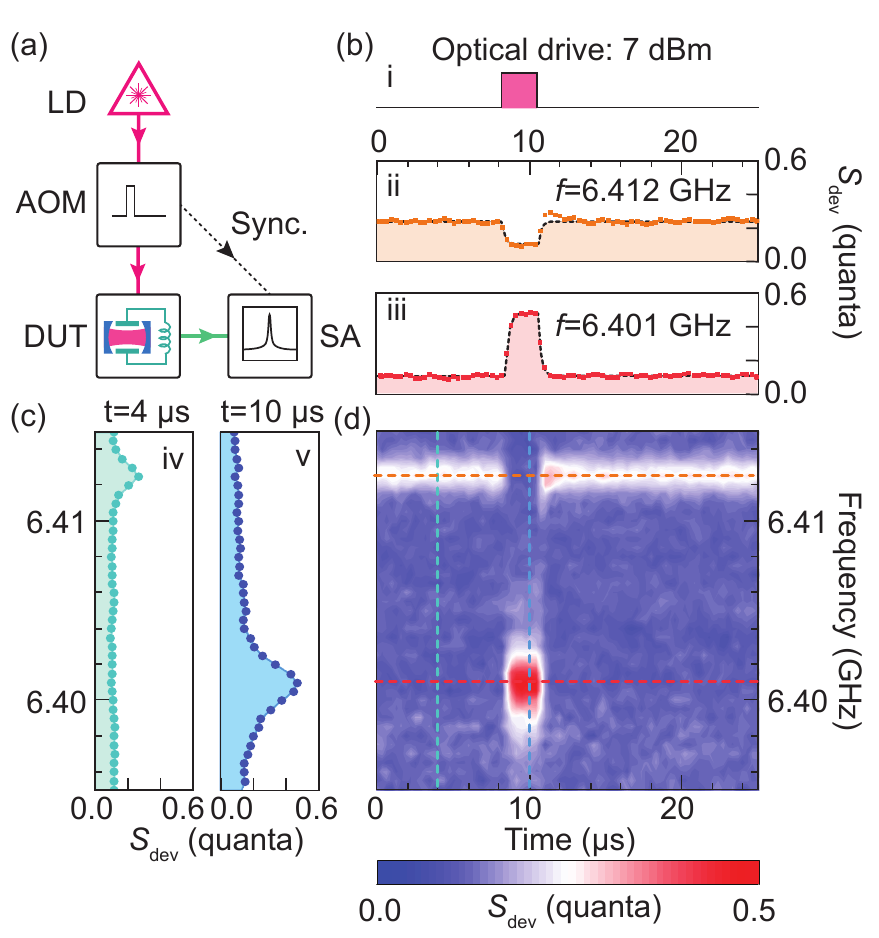}
\caption{(a) Simplified diagram of the experiment setup. LD: Laser diode. AOM: acousto-optic modulator. DUT: device under test. SA: spectrum analyzer or a lock-in amplifier that measures the noise spectrum. A CW laser is modulated with a pulse signal via an AOM before being sent into the device. The microwave noise goes through a low-noise amplification chain and is recorded by a lock-in amplifier, which are together represented by a spectrum analyzer in the diagram. The heat map (d) shows the time evolution of the noise spectrum. The time traces of noise power spectral density at original resonance frequency (ii) and shifted frequency (iii) are shown in (b), compared to the illustrated optical drive power in (i). (c) shows the snapshots of power spectral density when the pulsed laser drive is off (iv) and on (v). Data in (b) and (c) are cross-sections of (d) marked in correspondingly-colored dashed lines.}
\label{fig2}
\end{figure}

To study the dynamical noise performance of the device, we implement a fast precision measurement to monitor the power spectral density of the microwave port output noise as illustrated in Fig.\,\ref{fig2}(a).
For better noise photon resolution, a traveling wave amplifier with GHz-level bandwidth is utilized to achieve a low added noise readout chain. To characterize the absolute gain and added noise of the output chain, a noise thermometry calibration was performed and revealed the output line added noise as around 4 quanta at the frequencies of interest \cite{supplemental}, which is calibrated out in the following results.
The noise power measurement is carried out in a lock-in amplifier (LIA) with a low-pass filter bandwidth of 700 kHz, which corresponds to a time constant $t_c=202\,\mathrm{ns}$.\par

Measurement results at 7\,dBm pulsed optical drive is shown in Fig.\,\ref{fig2}. The heat map (Fig.\,\ref{fig2}(d)) depicts the time evolution of the light-induced noise power spectrum. 
A consistent background noise of around 0.11 quanta is observed across the entire frequency range of interest, which is more clear in Fig.\,\ref{fig2}(b) and (c).
During the light-off period, a peak on the noise spectrum (Fig.\,\ref{fig2}(c)\romannum{4}) at the microwave resonant frequency 6.412 GHz indicates a finite thermal excitation. 
In comparison, during the light-on period, as shown in Fig.\,\ref{fig2}(c)\romannum{5}, the superconducting resonant frequency down-shifts 11\,MHz, and the resonance linewidth is broadened. Moreover, the noise peak rises, indicating a stronger thermal excitation of the microwave mode.
The temporal evolution of the noise power spectral density at these two center frequencies are captured in Fig.\,\ref{fig2}(b). 
In light-off period, Fig.\,\ref{fig2}(b)\romannum{2}\ suggests that the microwave thermal excitation is relatively stable, it could be attributed to a thermal bath with a large time constant. 
From Fig.\,\ref{fig2}(b)\romannum{3}, it is evident that during the light-on period, a new thermal equilibrium is established.
The rising and falling edges of the measured transition agree well with the theoretical exponential functions (dashed curve), of which the time constant is equivalent to the measurement time constant $t_c=202\,\mathrm{ns}$.
This suggests that the actual transition time, which includes the resonance frequency shift as well as the heating when the light is turned on should be much shorter than 200\,ns. 
\par

\begin{figure*}[ht]
\begin{centering}
\includegraphics{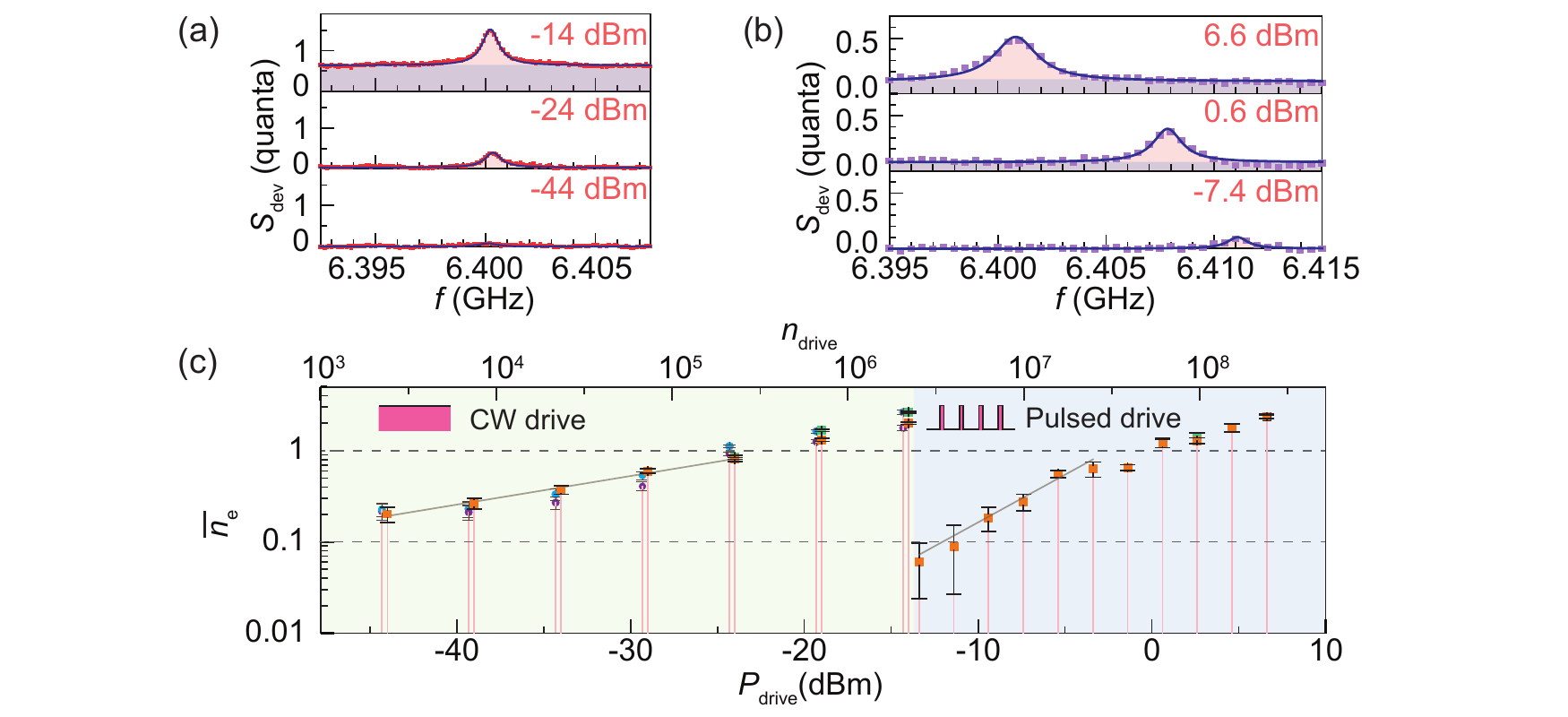}
\par \end{centering}
\caption{
(a) and (b) shows noise power spectra $S_\mathrm{dev}$ when CW and pulsed drives of different power are applied. 
The solid curve is the theoretical fit with coupling parameters independently characterized. (c) summarizes the microwave mode thermal occupancy $\bar{n}_\mathrm{e}$ at different optical drives. The x-axis represents the power of the CW drive or the peak power of the pulsed drive. Green data points, if visible, represent the upper bound of the occupancy given the uncertain nature of the observed noise background in the spectrum. Accordingly, the orange data points represents the lower bound of the occupancy. CW results from a different cooldown are shown as the purple (lower bound) and blue (upper bound) data points. }
\label{fig3}
\end{figure*}

The fast frequency shift and linewidth broadening when light is shined on the superconducting material are typically attributed to the superconductor absorption of photons \cite{2000_apl_quasiparticles_NbN,Nature_2003_SC_photondetector_mazin,PRL2011NbNquasipartical,APL2013TiNquasipartical}. During this process, the photon absorption breaks cooper pairs, generates quasi-particles, and consequently increases the kinetic inductance and induces extra dissipation in the superconductor.
Our results imply that for NbN the quasi-particle recombination time is much faster than 200\,ns at 40\,mK regime \cite{PRL2011NbNquasipartical,APL2013TiNquasipartical}.

From the microwave output noise power spectra, we can infer the optically induced effective thermal bath occupancy as well as microwave mode thermal excitations. 
As introduced in detail in reference \cite{RadiativeCoolingXu2020}, the output noise spectrum can be writtern as:
\begin{equation}
\label{psd_theory}
   S_\mathrm{dev}(\omega)=\mathcal{R}(\omega)\bar{n}_\mathrm{ex}+\big(1-\mathcal{R}(\omega)\big)\bar{n}_\mathrm{en}+\Delta\bar{n}_\mathrm{out,add},
\end{equation}
in which $\mathcal{R}(\omega)=1-\kappa_\mathrm{b,in}\kappa_\mathrm{b,ex}/\big({(\kappa_\mathrm{b}/2)^2+(\omega-\omega_\mathrm{0})^2}\big)$.
Here the first and second term are contribution from the external bath and intrinsic bath, respectively, and $\Delta\bar{n}_\mathrm{out,add}$ represents the additional output line added noise induced by light. 
$\kappa_\mathrm{b,in}$ and $\kappa_\mathrm{b,ex}$ are the coupling rates to the intrinsic and the external decay channels, and $\omega_\mathrm{0}$ is the angular resonant frequency of the resonator.
$\bar{n}_\mathrm{en}$ and $\bar{n}_\mathrm{ex}$ are the respective bath occupancy of the intrinsic bath and the external bath. 
 
By fitting each noise spectrum to Eq.\,\ref{psd_theory}, the intrinsic and external thermal bath occupancy can be found. 
Given both bath occupancies, the microwave mode thermal excitation can be calculated as the weighted average of the two thermal baths
\begin{equation}
\label{occupancy}
   \bar{n}_\mathrm{e}=\frac{\kappa_{\mathrm{b,in}}\bar{n}_\mathrm{en}+\kappa_{\mathrm{b,ex}}\bar{n}_\mathrm{ex}}{\kappa_{\mathrm{b,in}}+\kappa_{\mathrm{b,ex}}}.
\end{equation}

Fig. \ref{fig3}(b) shows microwave noise power spectra in the light-on period of different pulsed drives. 
In comparison, noise power spectra with different CW optical drive are plotted in Fig.\,\ref{fig3}(a).
Solid lines in both Fig.\,\ref{fig3}(a) and (b) are the theoretical fit to Eq.\,\ref{psd_theory} with the coupling parameters ($\kappa_\mathrm{ex}$ and $\kappa_\mathrm{in}$) each independently evaluated from the reflection spectrum by probing with a weak coherent tone.
Significant frequency shifts of the microwave resonance is observed when a strong pulsed drive is applied due to optical absorption by superconductor. 
It is worth noting that, a weak background noise (shaded in grey) is observed in power spectra when a strong CW or pulsed optical drive is applied. 
In the pulsed measurement, the background noise power stays consistent during the entire measurement time span, indicating its large time constant compared to the pulse period $1\,\mathrm{ms}$. Further investigation is needed to identify the source of the excess background noise. It could be attributed to the heated external bath of the superconducting resonator, or increased added noise of the amplification chain due to the optical excitation.

Based on the fitted microwave noise power spectra, the average thermal occupancy of the superconducting mode when driven with different optical pump is plotted in Fig. \ref{fig3}(c). Due to the uncertain nature of the background noise that appears in some higher power CW and pulsed measurements, lower bound (orange/purple) and upper bound (green/blue) of the microwave mode occupancy are provided. 
The upper bound values are obtained by considering the background noise as the heated external bath $\bar{n}_{ex}$ of the superconducting resonator, which contributes to the microwave thermal occupancy; the lower bound corresponds to the case when the background noise is considered as additional added noise in the output chain $\Delta\bar{n}_\mathrm{out,add}$, which does not contribute to the microwave thermal excitation. 
Fitting errors are also marked on each data point. 
It can be seen that when applying pulsed drive of $-11.4\,\mathrm{dBm}$ peak power in the waveguide, the microwave mode occupancy $\bar{n}_\mathrm{e}=0.09\pm0.06$, corresponding to a mode which is at its quantum ground state with $92\pm5\%$ probability. 
When increasing the pulsed drive peak power, the thermal occupancy scales approximately in proportion (power=1.03) to the pulsed drive power.
Compared to the CW drive, the pulsed optical drive allows a larger peak power to boost the electro-optical coupling rate by around 3 orders of magnitudes, while maintaining close-to-ground state microwave resonance. The different scaling trend of thermal excitation to power between CW and pulsed operation suggests different heating mechanisms. 

\begin{figure}[h]
\includegraphics{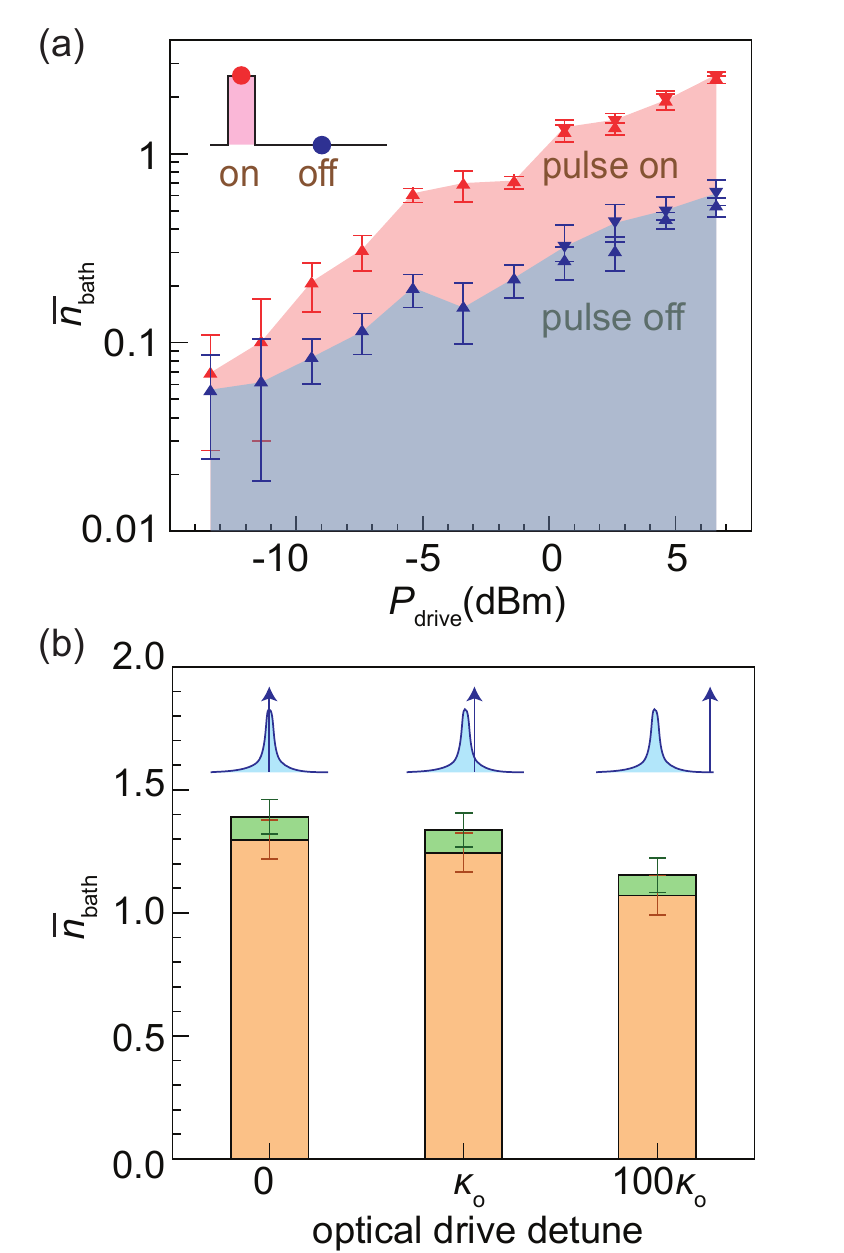}
\caption{(a) Steady state effective intrinsic bath occupancy when the pulsed drives are on and off duty. Lower bound and upper bound of some data points are provided as the triangle pointing up and down, respectively, due to the presence of the background noise of which origin is not known to us yet. (b) microwave intrinsic bath occupancy when applying 0.6\,dBm optical drives that are on-resonance, detuned by one linewidth and completely detuned. Lower bound (orange) and upper bound (green) are provided.
}
\label{fig4}
\end{figure}

\textit{Heating mechanism analysis.---}
To gain insight on the underlying heating mechanisms, we further analyze the time evolution of the noise spectrum with various optical drive powers.
Figure \ref{fig2} suggests two types of effective intrinsic baths with distinct time constants. The first is a slow-responding effective bath, which is responsible for the microwave noise during the light-off period. With no discernible change in slow bath occupancy for the entire pulse period, we can conclude that its time constant is much larger than the period of the pulsed drive $T=1\,\mathrm{ms}$. The second type is a fast-responding thermal bath with a ramp-up time much quicker than our data acquisition time constant 202\,ns, which contributes to the extra noise in the microwave resonator when the light is on.
The slow-responding bath can be attributed to heating due to optical absorption by dielectrics and other materials that stray light shines on. 
The fast-responding effective bath, on the other hand, likely originates from quasi-particles generation and recombination due to superconductor absorption of optical photons.
Another possible contribution to the fast thermal bath is localized dielectric heating in the optical cavity.
Compared to the slow bath, the fast bath is harder to manage in an electro-optical system with a strong optical drive, because it can not be suppressed by optimizing the pulsed drive configuration.
Figure \ref{fig4}(a) plots the microwave effective bath occupancy of the steady states when the optical drive is on and off.
According to the analysis above, the blue area represents the slow-responding bath occupancy and the the pink area represents the contribution from the fast-responding effective bath.
The result shows that the contribution of the fast-responding effective bath is initially smaller than that of the slow-responding bath with the lowest pulsed drive power, but is accountable for the majority of microwave noise in the higher pulsed drive power regime as it scales faster with the drive power. 

The next subject we study is the contribution to the microwave thermal excitation from different optical photon sources, among which we mainly focus on two categories: the cavity photons that are dissipated in the microring and those stray photons scattered off the chip-fiber interface.
For this device under test, based on the calibration at room temperature, 50\% of light in the fiber is not coupled to the waveguide and scattered off the chip-fiber interface. The rest is coupled to the waveguide and mostly dissipated in the microring for a resonant drive because of the critical-coupling condition, or, for a detuned drive, transmitted in low loss waveguide and then coupled back to the fiber.
To discriminate the impact of the optical photons scattered off the chip-fiber interface, we characterize microwave noise when applying a non-resonant optical drive.
In contrast to a resonant drive, a non-resonant drive does not launch cavity photons nor generate local heat in the microring.
Figure \ref{fig4}(b) shows light-on microwave intrinsic effective bath occupancy with 0.6\,dBm pulsed drives that are resonant, detuned by one-linewidth and completely detuned.
The difference in thermal bath occupancy when applying the resonant and the detuned drive is $17\pm7\%$.
This results suggest that despite the large mode overlap between the the microwave and the optical modes, optical cavity photons, which introduce dielectric heating as well as local radiation absorbed by superconductor, only contribute approximately 17\% of the microwave thermal excitation.
The leading factor of the microwave mode heating is thus concluded to be superconductor absorption of stray light which are scattered off the chip-fiber interface. 

\textit{Discussion and perspectives.---}
Admittedly, for the very device studied in this work, even with reduced microwave noise using the pulsed drive scheme, practical quantum applications still remains elusive.
One major limitation is the relatively weak coupling, which leads to insufficient conversion efficiency between the microwave mode and optical mode, because of the undesired gap between the superconducting and photonic chip (estimated to be 14\,$\mathrm{\mu m}$, see \cite{supplemental}).
To achieve a larger vacuum electro-optical coupling coefficient, one can use better bonding technique or employ the single-chip approach \cite{eo8_fan2018superconducting}, which promises approximately 1 order of magnitude enhancement. Switching to material platforms with higher Pockels coeffient such as lithium niobite and barium titanate can potentially bring about orders of magnitude improvement \cite{LN_modulator,eo5_LN_amir_2020,eo6_LN_loncar_2020,BTO_modulator}. Applying stronger optical pumps could also parametrically enhance the electro-optical coupling strength, which may require improved device power handling capability beyond what is achieved here.

Regarding the optically induced microwave noise, our work revealed two types of thermal bath with distinct time constants. The slow responding bath, which is determined by the average optical drive power, could be further suppressed by using lower duty cycle at a given peak drive power. The fast responding bath, which is a more dominant factor at higher optical drive powers, is relatively insensitive to the change of pulsed drive configuration within one order of magnitude, because the new steady state is rapidly established within 200\,ns after the drive is turned on. Separate measurements with $T=16\,\mathrm{ms}$ and duty cycle of 0.2\% yields similar results to that with $T=1\,\mathrm{ms}$ and 0.2\% duty cycle. 
To further reduce the fast responding bath occupancy at a given optical pump power, possible solutions include spatially distancing the chip-fiber interface from the superconducting resonator, creating proper shield to block stray light from the chip-fiber interface, and providing better thermal dissipation channels to the superconducting material, for example immersing the device in superfluid helium \cite{eo8_fan2018superconducting}.
When adopting an over-coupled configuration for the microwave resonator, the microwave resonance could also be radiatively cooled by exchanging heat with a cold bath through the microwave waveguide \cite{RadiativeCoolingXu2020,radiative_cooling_wang2019quantum}.


For an arbitrary quantum transduction between microwave and optical states, it is required to have at least 50\% conversion efficiency while maintaining low microwave noise, which means approximately 3 orders of magnitude of improvements needs to be made in terms of the enhanced electro-optical coupling coefficient $g_\mathrm{o}$.
In principle, we can bypass these demanding requirements by using quantum teleportation protocol \cite{1998PRL_CV_teleportation} with the assistance of a classical communication channel, which relies on continuous variable entanglement generation between microwave and optical modes by pumping the higher frequency optical mode \cite{CV_teleportation_PRL_2012,rueda2019_entanglement_source,entanglement_cczhong}.
 Assuming perfect homodyne measurement, with our current noise measurement results and device parameters under optical drives of different powers, a continuous variable teleportation fidelity exceeding the ``non-cloning threshold'' $2/3$ \cite{TeleportationNonCloningThreshold_pra_2001} is possible if the electro-optical coupling rates $g_\mathrm{o}$ improves 2 orders of magnitude \cite{supplemental}. With the rapid development of cavity electro-optical systems based on different material platforms, these goals seem promising \cite{eo8_fan2018superconducting,eo7_groundstate_fink_2020,eo5_LN_amir_2020,eo6_LN_loncar_2020}. Reduced optically induced microwave noise could also help lower the requirement for $g_\mathrm{o}$.

To summarize, we present a focused study of near-ground state microwave thermal excitations in a cavity electro-optical system driven by strong pulsed optical drives. 
Dynamical noise characterization reveals that the microwave noise is limited by a fast responding effective bath caused by superconductor absorption of stray drive photons scattered from the chip-fiber interface.
We show that with practical improvements, microwave-to-optical quantum interface based on pulsed cavity-electro optical system is highly feasible.


\noindent \textbf{Acknowledgments}\\
This work is supported by ARO grant W911NF-18-1-0020. The TWPAs used in this experiment are provided by IARPA and MIT Lincoln Laboratory. The authors also acknowledge partial supports from NSF (EFMA-1640959) and the Packard Foundation. The authors thank Y. Sun, S. Rinehart, K. Woods, and M. Rooks for assistance in device fabrication. 

\noindent \textbf{Author contributions}\\
All authors contributed to the manuscript. H.X.T. supervised the work.

\noindent \textbf{Additional information}\\
Supplementary Information accompanies this paper.

\noindent \textbf{Competing interests:} The authors declare no competing interests.

\end{document}